\documentclass[12pt]{article}
\usepackage{graphicx}

\begin{document}

\begin{center}
\Large {\bf Proof of the Holographic Formula
for Entanglement Entropy}
\end{center}

\bigskip
\bigskip

\begin{center}
Dmitri V. Fursaev
\end{center}

\bigskip
\bigskip

\begin{center}
{\it Dubna International University \\
  and\\
  the University Centre,\\
  Joint Institute for Nuclear Research\\
  141 980, Dubna, Moscow Region, Russia\\}
 \medskip
\end{center}
\begin{center}
  {\rm E-mail: \texttt{fursaev@theor.jinr.ru}}\\
  \medskip
\end{center}

\bigskip
\bigskip

\begin{abstract}
Entanglement entropy for a spatial partition of a quantum system is studied
in theories which admit a dual description in terms of the anti-de Sitter (AdS)
gravity one dimension higher.
A general proof of the
holographic formula which relates the entropy to the
area of a codimension 2 minimal hypersurface embedded in the bulk AdS space is given. The entanglement
entropy is determined by a partition function which is defined as a path integral
over Riemannian AdS geometries with non-trivial boundary conditions.
The topology of the Riemannian spaces puts restrictions on the choice of the minimal hypersurface
for a given boundary conditions.
The entanglement entropy is also considered in Randall-Sundrum  braneworld  models where
its asymptotic expansion is derived when the curvature radius of the brane is much larger than
the AdS radius. Special attention is payed to the geometrical structure of anomalous terms
in the entropy in four dimensions. Modification of the holographic formula by the higher
curvature terms in the bulk is briefly discussed.
\end{abstract}

\newpage

\section{Introduction}

Entanglement entropy measures  a degree of the  correlation between different subsystems
of a quantum system. In many-body theories the quantum entanglement is usually determined
when subsystems are spatially separated. A considerable attention has been paid in last few years
to studying entanglement entropy in 2D spin-chain models near the critical point
with the purpose to understand better collective effects in strongly correlated systems
(for a review of some of these results see \cite{Rico}).

In a similar way, one could pose the question about entanglement on the fundamental level,
in a quantum gravity theory.
It was conjectured in \cite{DF:06a}
that for partition of the ground state by a plane the entanglement
entropy per unit area of the
plane is
\begin{equation}\label{1.1}
{\cal S} = {1 \over 4G_N}~~,
\end{equation}
where $G_N$ is the gravitational
constant in the low-energy gravity sector of the theory.  ${\cal S}$ measures entanglement
of all genuine microscopical degrees of freedom of the fundamental theory.

Relation (\ref{1.1}) has been motivated by the black hole
physics\footnote{First indications on
a possible relation between quantum
entanglement and gravity phenomena should be  attributed to authors of \cite{Sr:93},\cite{BKLS}
who suggested to interpret the Bekenstein-Hawking entropy as the entanglement entropy.}:
the density of the
Bekenstein-Hawking entropy per unit area of the black hole horizon
is universal and equals $1 /(4G_N)$.
The fact that quantum entanglement
and gravity are connected phenomena is interesting
because one can study properties of the gravitational
couplings in "gravity
analogs", by using entanglement entropy in condensed matter systems \cite{DF:06a}.

If the fundamental theory lives in a space-time with number of dimensions higher than four
there is an issue of higher-dimensional extension of  (\ref{1.1}).
There may be at least two options. In theories of the Kaluza-Klein (KK) type all degrees of freedom
propagate in the bulk. In this case partition of all KK modes by a hypersurface $\cal B$ in
four-dimensions is equivalent to partition of higher-dimensional fields by the surface
$\tilde{\cal B}={\cal B} \times \Omega_{D-4}$ where $\Omega_{D-4}$ is a small compact space
of $D-4$ extra-dimensions. If (\ref{1.1}) is true
the entanglement entropy should be
\begin{equation}\label{1.2}
S= {{\cal A} \over 4G_N}= {{\tilde{\cal A}} \over 4G_N^{(D)}}~~~,
\end{equation}
where ${\cal A}$ is the area of ${\cal B}$ and
${\tilde{\cal A}}={\cal A}\cdot {\mbox{vol}~ \Omega_{D-4}}$ is the volume of
$\tilde{\cal B}$. Equation (\ref{1.2}) takes into account the relation
$G_N^{(D)}=G_N \cdot {\mbox{vol}~ \Omega_{D-4}}$, between higher-dimensional,
$G_N^{(D)}$, and low-dimensional, $G_N$, gravitational couplings.

The other option is the theories with large extra dimensions. In these theories the gravity
remains higher-dimensional while matter fields are confined on a brane and do not propagate in the
bulk.
For this reason, the way how the separating surface is extended in extra-dimensions
should be determined by the bulk gravity equations.

There is a particular example where this issue can be investigated:
the Randall-Sundrum model where the bulk is the anti-de Sitter (AdS) gravity. If $\cal B$ is
a separating hypersurface in the brane theory, its extension $\tilde{\cal B}$ has to be a minimal
surface embedded in the AdS space. The boundary of
$\tilde{\cal B}$  on the brane is $\cal B$.
With this prescription the entropy in the brane theory takes the form (\ref{1.2}),
\begin{equation}\label{1.3}
S= {{\tilde{\cal A}} \over 4G_N^{(d+1)}}~~~,
\end{equation}
where $G_N^{(d+1)}$ is the gravitational constant in the bulk and $d$ is the number of dimensions
on the brane.

Equation (\ref{1.3}) is the "holographic formula" which was first suggested by Ryu and
Takayanagi \cite{RT:06a}, \cite{RT:06b}
for the entropy of conformal field theories (CFT) which admit a dual description
in terms of the AdS gravity one dimension higher. The remarkable property of the result of
\cite{RT:06a}, \cite{RT:06b} is that (\ref{1.3}) enables one to study quantum entanglement
in strongly correlated systems, such as strongly coupled gauge theories, where the direct field theoretical (QFT) computations
for the entropy are extremely difficult.

Ryu and Takayanagi presented an intuitive derivation
of (\ref{1.3}) by using AdS/CFT correspondence \cite{Ma}--\cite{GKP}
and provided some examples in its support.
The analysis of  \cite{RT:06a}, \cite{RT:06b}
was restricted by static problems
and $\tilde{\cal B}$ was assumed to be a static minimal surface in a Lorenzian AdS background.
In \cite{Em:06} equation (\ref{1.3}) was used to argue that
the entropy of
a black hole in a braneworld model in three dimensions is an entanglement entropy.
In \cite{IKSY} implications of (\ref{1.3}) were considered  for the entropy
of a de Sitter brane.

The main purpose of our work is to understand better where and at
which conditions relation (\ref{1.3}) comes from.
The present paper is organized as follows. The proof of  the holographic formula
is given in Section 2. For a finite-temperature theory the entanglement entropy can be derived in a statistical
mechanical manner from a partition function $Z^{CFT}(\beta,T)$ of the boundary CFT.
$Z^{CFT}(\beta,T)$
is defined on a class of
Riemannian manifolds ${\cal M}_n$ (manifolds with the Euclidean signature).
The parameter $\beta=2\pi n$, where $n$ is a natural number,
can be interpreted as an inverse "temperature". By using the AdS/CFT conjecture we write
$Z^{CFT}(\beta,T)$ as
a path integral in the bulk AdS gravity with the boundary condition that the conformal
boundary of "hystories" $\tilde{\cal M}_n$ involved
in the path integral  belongs to the conformal class of ${\cal M}_n$.
The bulk spaces have conical singularities on codimension 2 hypersurfaces $\tilde{\cal B}$
which are "holographically dual" to the separating surface $\cal B$ in the boundary CFT.
Evaluation of the path integral in the saddle point approximation picks up a space
$\tilde{\cal M}_n$ where the classical AdS gravity action has an extremum. The variational
principle for the action implies that $\tilde{\cal B}$ has to be a minimal surface embedded in
a Riemannian AdS background.
It then immediately follows that in the saddle point approximation the entanglement entropy
is given by (\ref{1.3}). The derivation of (\ref{1.3}) answers some questions
left open in \cite{RT:06a}, \cite{RT:06b} and makes the statement more precise.
For instance, the topology of a Riemannian space puts restrictions on the choice of
the minimal hypersurface under chosen boundary conditions.
Connection of  the entanglement entropy with the path integral in AdS gravity enables
one to study entanglement in the theories which are stationary but
not static and to
consider theories with different phases.

In Section 3 we study consequences
of (\ref{1.3}) in RS brane-world scenario. In particular, we derive the asymptotic geometrical
structure of the entropy when the curvature radius of the brane is much larger than
the AdS radius. In Section 4
a general form of the "anomalous term" in the entropy is given
in four-dimensional brane theories, including the case when $\cal B$ has non-vanishing
extrinsic curvatures.
Modification of the holographic formula by the higher curvature terms in the bulk
action is briefly analyzed in Section 5.
We discuss our results in Section 6. It is emphasized that
if the corrections caused by the curvature of the brane are neglected
the area density of the ground state entanglement entropy in the theory on the brane
is given by  (\ref{1.1}) where $G_N$ is the Newton coupling in the gravity induced
on the brane, in complete agreement with the suggestion of \cite{DF:06a}.

\section{Holographic formula for the entanglement entropy}
\setcounter{equation}0

An extended form of the AdS/CFT conjecture \cite{Ma}--\cite{GKP}, which we follow in this paper, asserts
that a supergravity theory in
the anti-de Sitter space is dual to a conformal field theory on the boundary
of that region.
The correspondence takes the following form\footnote{We consider Euclidean theory.
We also assume that the bulk space has $d+1$ dimensions. All geometrical structures in the
bulk  are denoted with a tilde. The corresponding "dual" quantities on the boundary will be denoted
by the same letters without the tilde. For example, the bulk Riemannian manifold
$\tilde{\cal M}$ corresponds to a boundary manifold $\cal M$.}
\begin{equation}\label{2.1}
Z^{CFT}[h]=\int [Dg]\exp(-I_{\mbox{gr}}[g])~~~.
\end{equation}
Equation (\ref{2.1}) relates the partition function $Z^{CFT}[h]$ of a CFT
($\ln Z^{CFT}[h]$ is a generating functional of connected Green's functions)
on a manifold ${\cal M}$
with a metric $h_{\mu\nu}$ to the gravity theory in a space-time one dimension higher.
The path integral in the right hand side (r.h.s.) in (\ref{2.1}) is taken over all geometries
which induce a given conformal class of metrics $h_{\mu\nu}$ at their conformal boundary.
We denote bulk manifolds  and metrics as $\tilde{\cal M}$ and $g_{KL}$,
respectively.
The action in (\ref{2.1}) is the gravity action
\begin{equation}\label{2.2}
I_{\mbox{gr}}=-{1 \over 16\pi G_N^{(d+1)}}\int_{\tilde{\cal M}} d^{d+1}x\sqrt{g}
\left(R+{d(d-1) \over l^2}\right)
-{1 \over 8\pi G_N^{(d+1)}}\int_{\partial \tilde{\cal M}} d^{d}x\sqrt{h}K~~~
\end{equation}
with the negative cosmological constant
$\Lambda=-d(d-1)/(2l^2)$.
To avoid the volume divergences of $I_{\mbox{gr}}$ one considers a cut of the bulk
manifold at some $d$-dimensional hypersurface $\partial \tilde{\cal M}$ at some
large radius. In the limit of infinite radius $\partial \tilde{\cal M}$
belongs to the conformal class of ${\cal M}$.
The boundary term in (\ref{2.2}), which depends on the metric $h_{\mu\nu}$ induced on
$\partial \tilde{\cal M}$ and on the trace of its extrinsic curvature $K$, is the
Gibbons-Hawking term
which is necessary for a well-defined variational problem.
The corresponding conformal theory on $\partial \tilde{\cal M}$
has an ultraviolet cutoff related to the radius of the boundary.

In analogy with (\ref{2.1}) we formulate the method of computing entanglement entropy
$S$
in the boundary CFT in terms of the path integral over AdS metrics.
We start with a recipe of computing $S$ in a QFT described in detail in \cite{DF:06a}.

Consider a QFT model of a field variable $\phi$ defined on a
$(d-1)$ dimensional hypersurface $\Sigma$.
The state of the system,
can be described in general by the density matrix $\hat{\rho}$ and in what follows we
assume that it is a state at temperature $T$. The ground state can be obtained in the limit
of zero temperature. When applying our results to conformal theories which have AdS duals,
$\Sigma$ can be, for example, a hyperplane $R^{d-1}$, a hypersphere $S^{d-1}$, or a Lobachevsky
hyperbolic space $H^{d-1}$, see \cite{ct1}. In these cases
${\cal M}$ is $R^{d-1} \times S^1$, $S^{d-1}\times S^1$, or $H^{d-1} \times S^1$. There may be
more complicated situations which we mention below.

Suppose that $\Sigma$ is divided onto two parts, $\Sigma_1$ and $\Sigma_2$ by a
codimension 2 hypersurface
$\cal B$ and define a reduced density matrix for the region $\Sigma_1$,
$$
\hat{{\rho}}_1=\mbox{Tr}_2 \hat{\rho}~~,
$$
by taking trace over the states located in the region $\Sigma_2$.
The entanglement entropy
is defined as the von Neumann entropy
\begin{equation}\label{2.3}
S_1=-\mbox{Tr}_1 \hat{{\rho}}_1\ln \hat{{\rho}}_1~~.
\end{equation}

\begin{figure}[h]
\begin{center}
\includegraphics[height=6.5cm,width=7.5cm]{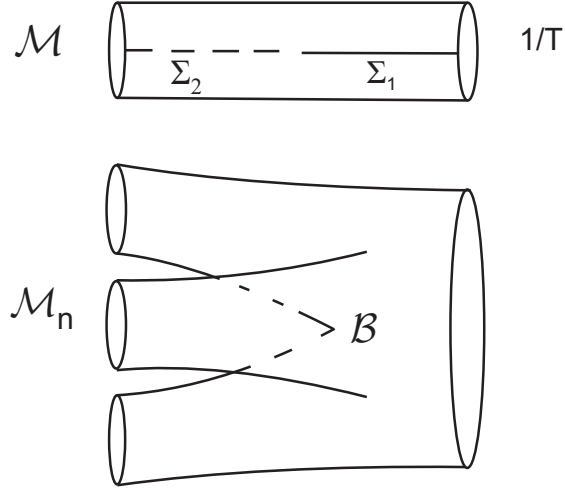}
\caption{\small{Spaces ${\cal M}$ and ${\cal M}_n$ (for $n=3$) are shown
schematically on upper and lower pictures. The theory at temperature $T$ is defined on
a space $\Sigma_1\cup\Sigma_2$ shown as an interval.}}
\label{f1}
\end{center}
\end{figure}

Analogously, one can define the entanglement entropy $S_2$ in the region $\Sigma_2$ by tracing
the density matrix over the states located in the region $\Sigma_1$.
If the system is in a pure state $|\psi\rangle$, i.e. $\hat{{\rho}}=|\psi\rangle\langle\psi |$,
it is not difficult to show that $S_1=S_2$, see \cite{Sr:93}.
In what follows we do all our computations for the entropy in $\Sigma_1$.
To simplify the notations we write the entropy and all associated objects without
subscript 1. Distinctions in the entropies for different regions will be explained below.

Let us introduce a discrete parameter $\beta=2\pi n$,
and define a "partition function"
\begin{equation}\label{2.4}
Z(\beta,T)=\int [D\tilde{\phi}]~e^{-I[\phi,\beta,T]}~~~,
\end{equation}
where $I[\phi,\beta,T]$ is the classical action for the field $\phi$ defined on
a Riemannian manifold ${\cal M}_n$.

The definition of ${\cal M}_n$ and $Z(\beta,T)$
is as follows. If $n=1$ (\ref{2.4}) coincides with the partition function of the given
field $\phi$ in space $\Sigma$
at finite temperature $T$.
The background manifold in this case is ${\cal M}_1=\Sigma \times S^1\equiv{\cal M}$, where
$S^1$ is the circle of the length $T^{-1}$. To obtain ${\cal M}_n$
for $n>1$ one has to take $n$ copies of ${\cal M}$ each with a cut along $\Sigma_1$ and
glue them along the cuts as is shown on Fig. \ref{f1}.
The entanglement entropy  is given by the
equation \cite{DF:06a}
\begin{equation}\label{2.5}
S(T)=-\lim_{\beta \rightarrow 2\pi}~ \left(\beta {\partial \over
\partial \beta}-1\right) \ln Z(\beta,T)~~~.
\end{equation}
The operation with the parameter $\beta$ in (\ref{2.5}) should be understood in the following way:
one first computes $Z(\beta,T)$ for $\beta=2\pi n$, and then replaces
$\beta$ with a continuous parameter. This can be done
even if ${\cal M}_n$ itself cannot be defined at arbitrary $\beta$.
It will be important for us that (\ref{2.5}) coincides with the formula for the
entropy in statistical mechanics if $Z(\beta,T)$
is interpreted as a partition function and $\beta^{-1}$ as a
temperature\footnote{The parameter $\beta^{-1}$ should not be confused with the physical
temperature $T$.}.

In the same way, by cutting $n$ copies of ${\cal M}$ and gluing along $\Sigma_2$ one can obtain
the entanglement entropy in the region $\Sigma_2$.

Let us emphasize that definition (\ref{2.5}) can be also applied to the entropy in theories
which are stationary but not static. To give an example, suppose that $\Sigma$ has an
axial isometry generated by a Killing vector field $\partial_\varphi$. One can study the
entanglement entropy in the frame of reference which rigidly rotates with an angular
velocity $\dot{\varphi}=\Omega$.  The space $\cal M$ is a torus which
is obtained by identifications $(\tau,\varphi)\sim (\tau,\varphi+2\pi)$ and
$(\tau,\varphi)\sim (\tau+1/T,\varphi+\Omega/T)$ (i.e., along the trajectories generated
by $\partial_\varphi$ and $\partial_\tau+\Omega \partial_\varphi$). Gluing $n$ copies of
$\cal M$  along the cut $\Sigma_1$ yields ${\cal M}_n$.
Computation of the entanglement entropy
in this setting has to be accompanied by the change $\Omega$ to $i\Omega$ in the final results.

Equation (\ref{2.5})  is a convenient starting point for computing the entropy in a
CFT by using the bulk AdS gravity. To this aim we treat $Z^{CFT}(\beta,T)$
as a genuine partition function but taken for a particular class of background spaces.
Then relation (\ref{2.1}) enables one to represent it
as
\begin{equation}\label{2.6}
Z^{CFT}(\beta,T)=\int [Dg]\exp(-I_{\mbox{gr}}[g])~~~.
\end{equation}
The path integral in the right hand side in (\ref{2.6}) is taken over all $(d+1)$
dimensional geometries $\tilde{\cal M}_n$
whose asymptotic boundary is related to ${\cal M}_n$ by a
conformal transformation. Like in (\ref{2.1}) formula (\ref{2.6})
implies a spatial cutoff in the bulk.

Description of "hystories" in (\ref{2.6}) is parallel to the construction
of ${\cal M}_n$.
First, one considers all AdS geometries whose asymptotic
boundaries are from the conformal class of ${\cal M}$. We denote these geometries
$\tilde{\cal M}$. Then, one makes different cuts of $\tilde{\cal M}$ along
$d$--dimensional hypersurfaces $\tilde{\Sigma}_1$. One requires that the
asymptotic boundary of $\tilde{\Sigma}_1$ is conformal to the cut $\Sigma_1$ in ${\cal M}$.
This boundary condition does not fix $\tilde{\Sigma}_1$ uniquely.
There may be different cuts of $\tilde{\cal M}$ which have this property.
Finally, by taking $n$ identical copies of $\tilde{\cal M}$ with the same cut
$\tilde{\Sigma}_1$
and gluing them along the cuts one gets a space $\tilde{\cal M}_n$ with the
required boundary
condition.

Integral (\ref{2.6}) can be computed in the saddle point approximation.
To do the computation one has to take into account
that $\tilde{\cal M}_n$ have conical singularities which lie on
$(d-1)$-dimensional hypersurfaces $\tilde{\cal B}$ where all $n$ cuts meet. The surplus
of the conical angle  around each point of $\tilde{\cal B}$ is $\beta-2\pi=2\pi(n-1)$.
The classical action (\ref{2.2}) in this case should be
$$
I_{\mbox{gr}}=-{1 \over 16\pi G_N^{(d+1)}}
\int_{\tilde{\cal M}_n/\tilde{\cal B}} d^{d+1}x\sqrt{g}\left(R+{d(d-1) \over l^2}\right)
-{1 \over 8\pi G_N^{(d+1)}}\int_{\partial \tilde{\cal M}_n} d^{d}x\sqrt{h}K
$$
\begin{equation}\label{2.7}
-{1 \over 8\pi G_N^{(d+1)}}(2\pi-\beta)\tilde{\cal A}~~~.
\end{equation}
The integral in the r.h.s. of (\ref{2.7}) goes over the regular region
$\tilde{\cal M}_n/\tilde{\cal B}$ of $\tilde{\cal M}_n$.
The delta-function like singularities in the curvature $R$ at $\tilde{\cal B}$
result in the last term in the r.h.s. of (\ref{2.7}) where
$\tilde{\cal A}$ is the volume of $\tilde{\cal B}$.

As in (\ref{2.2})  one considers a cut of the bulk
manifold at some $d$-dimensional hypersurface $\partial \tilde{\cal M}_n$
to avoid the volume divergences. The space $\partial \tilde{\cal M}_n$ has
conical singularities and one might worry that they yield additional boundary terms.
The structure
of  possible boundary terms could be
$$
{1 \over G_N^{(d+1)}}\int_{\partial \tilde{\cal B}} d^{d-1}x\sqrt{\sigma}f(\beta,{\cal R}, {\cal K},...)
$$
where the integral goes over $(d-1)$-dimensional hypersurface $\partial \tilde{\cal B}$,
a location of singular points on $\partial \tilde{\cal M}$,
with the determinant of the metric $\sigma$.
Function $f(\beta,{\cal R}, {\cal K},...)$ should be a local invariant of all possible
combinations of internal, ${\cal R}$, and external, ${\cal K}$, characteristics of
$\partial \tilde{\cal B}$. The problem is that $f$ has to have the dimension
of the length, while the
invariants have the dimension of the inverse length in some non-negative power.
Note also that one cannot use the AdS radius $l$ to construct combinations with required
dimensionality because the form of boundary terms (for off-shell metrics) does not depend
on the presence of the cosmological constant in the bulk. Thus, we conclude that
conical singularities do not generate boundary terms.

In the saddle point approximation (\ref{2.6}) yields
\begin{equation}\label{2.8}
\ln Z^{CFT}(\beta,T)\simeq -I_{\mbox{gr}}[\bar{g}]~~~,
\end{equation}
where the bulk metric $\bar{g}$ is a point where (\ref{2.7}) has an extremum,
\begin{equation}\label{2.9}
\delta I_{\mbox{gr}}[\bar{g}]=0~~~
\end{equation}
under the given boundary conditions. Equation (\ref{2.8}) holds in the case when there
is a dominating contribution from one of extremal configurations, so that other contributions
can be neglected. We will assume that this is the case.
Because one has to consider variational procedure
at a fixed value of $\beta$ one gets from (\ref{2.7}), (\ref{2.9})
two sorts of equations: the standard Einstein equations in the bulk and equation
for $\tilde{\cal B}$ which ensures that the variation of the last term in the r.h.s.
of (\ref{2.7}) vanishes,
\begin{equation}\label{2.10}
R_{KL}-\frac 12 \bar{g}_{KL} R+\Lambda \bar{g}_{KL}=0~~~,
\end{equation}
\begin{equation}\label{2.11}
\delta{\tilde{\cal A}}=0~~~.
\end{equation}
Condition (\ref{2.11}) results in the Nambu-Goto equations which describe embedding equations
of $\tilde{\cal B}$ as a minimal hypersurface
in bulk space $\tilde{\cal M}_n$.  The metric $\bar{g}$ of  $\tilde{\cal M}_n$
is a solution
to (\ref{2.10}). The boundary condition for (\ref{2.11})
requires that asymptotic conformal boundary
of $\tilde{\cal B}$ belongs to the conformal class of $\cal B$ (which is the separating
surface of $\Sigma$ and the place of location of conical singularities of ${\cal M}_n$).
It should be noted that spaces $\tilde{\cal M}_n$ and $\tilde{\cal M}$ are
identical outside the conical singularities. Therefore, $\tilde{\cal B}$ should be also
a minimal hypersurface in  $\tilde{\cal M}$.

One can use now (\ref{2.5}) and
(\ref{2.7}) to get holographic formula (\ref{1.3}) for the entanglement entropy
in a CFT theory
$$
S^{CFT}(T)=-\lim_{\beta \rightarrow 2\pi}~ \left(\beta {\partial \over
\partial \beta}-1\right) \ln Z^{CFT}(\beta,T)
$$
\begin{equation}\label{2.12}
\simeq {{\tilde{\cal A}} \over 4G_N^{(d+1)}}~~~.
\end{equation}

\begin{figure}[h]
\begin{center}
\includegraphics[height=5.5cm,width=7.5cm]{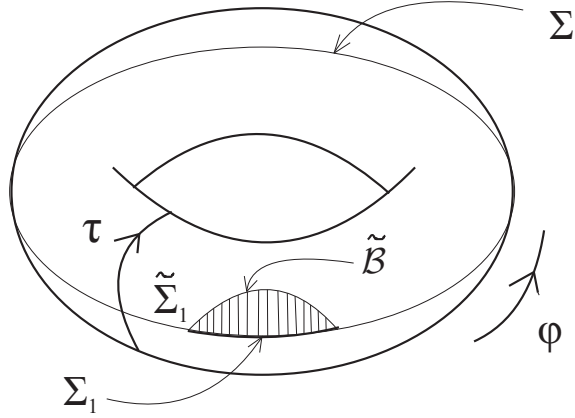}
\caption{\small{A Euclidean  BTZ black hole $\tilde{\cal M}$
is shown as a solid torus.
The shaded region is the hypersurface $\tilde{\Sigma}_1$ with the boundary $\Sigma_1$.
It lies inside the torus. The manifold $\tilde{\cal M}_n$ corresponding to the BTZ black
hole is obtained by cutting and gluing $n$ copies of $\tilde{\cal M}$ along $\tilde{\Sigma}_1$.}
}
\label{f2}
\end{center}
\end{figure}

Let us summarize conditions for (\ref{2.12}).

1) The formula holds in the semiclassical approximation when contribution
of the bulk gravitons (quantum fluctuations of the bulk geometry) is neglected.
See discussion of this point in Section 5.

2) A single dominating "trajectory", $\bar{g}$, in the
path integral (\ref{2.6}) which solves
bulk equations (\ref{2.10}) is implied.

3) $\bar{g}$ has the positive signature. Thus, the minimal surface $\tilde{\cal B}$ is
embedded in the Riemannian AdS background $\bar{g}$.
If the boundary space $\cal M$ is static $\tilde{\cal B}$ is also a minimal
static surface in the Lorenzian section of $\bar{g}$, the case initially
considered in \cite{RT:06a}, \cite{RT:06b}.

4) By the construction, the choice of $\tilde{\cal B}$ implies that there is a
cut $\tilde{\Sigma}_1$ inside the bulk space $\tilde{\cal M}$ such that the boundary
$\partial \tilde{\Sigma}_1$ consists of $\tilde{\cal B}$ and a conformal infinity
which belongs to the conformal class of $\Sigma_1$.  The boundary
$\partial \tilde{\Sigma}_1$ does not have other elements.

The last condition enables one to understand why in general  the entanglement
entropy in the region $\Sigma_1$ is different from the entanglement in its
completion $\Sigma_2$. This happens
when there are no cuts
in $\tilde{\cal M}$ connecting $\Sigma_2$ to $\tilde{\cal B}$.

The latter situation occurs  at finite temperatures  for topological reasons.
Let us illustrate it by using $AdS_3/CFT_2$.
If the $CFT_2$ is at non-zero temperature $T$ its gravity dual is a
BTZ black hole \cite{BTZ} given by the metric
$$
ds^2=(r^2-r_+^2)d\tau^2+{l^2 \over r^2-r_+^2} dr^2+r^2d\varphi^2=
$$
\begin{equation}\label{btz}
r_+^2\sinh^2(\rho /l)  d\tau^2+d\rho^2+r_+^2\cosh^2(\rho /l) d\varphi^2 ~~~,
\end{equation}
where $r=r_+\cosh(\rho /l)$, $0\leq \tau \leq 2\pi l/r_+$, $0\leq \varphi \leq 2\pi$.
The Euclidean horizon is located at $r=r_+$ or $\rho=0$. The Euclidean BTZ black hole has the
topology of a solid torus, as is shown in Fig. \ref{f2}.

\begin{figure}[h]
\begin{center}
\includegraphics[height=7.0cm,width=8.0cm]{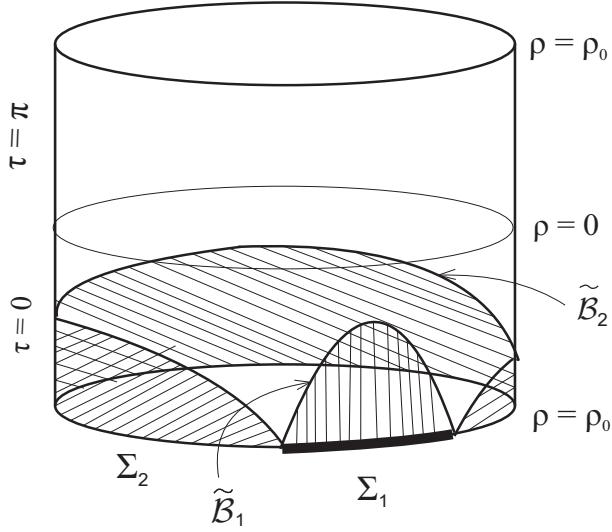}
\caption{\small{A slice by a plane of the torus shown on figure \ref{f2}.
The Euclidean horizon $\rho=0$ lies on the slice. The shaded regions are the hypersurfaces
$\tilde{\Sigma}_1$ and $\tilde{\Sigma}_2$. Their boundaries
inside the slices, $\tilde{\cal B}_1$ and $\tilde{\cal B}_2$, are the geodesic
lines.}}
\label{f3}
\end{center}
\end{figure}

To get rid off the volume divergences we assume that $0\leq \rho \leq \rho_0$, where $\rho_0$
is a cutoff. The surface $\Sigma$ is a circle $\rho=\rho_0$, $\tau=0$, while
$\Sigma_1$ is a segment of this circle as is shown on
Fig. \ref{f2}. It can be demonstrated that constant $\tau$ geodesics which start and end at the circle
$\rho=\rho_0$, $\tau=0$ make a turn before reaching the horizon $\rho=0$. One of such
geodesics, $\tilde{\cal B}_1$,  restricts the cut $\tilde{\Sigma}_1$ of the torus. The cut
starts at $\Sigma_1$ and lies in the half-plane $\tau=0$. The other geodesic line,
$\tilde{\cal B}_2$, which connects the ends of $\Sigma_1$ should be associated
to the completion $\Sigma_2$, see Fig. \ref{f3}.

The partition function $Z^{CFT}(\beta,T)$ of the CFT in the semiclassical
approximation is given by the classical action (\ref{2.8}) on a space $\tilde{\cal M}_n$
which is obtained by cutting along $\tilde{\Sigma}_1$ and gluing  $n$ copies of the
torus.

The cut of the torus $\tilde{\Sigma}_1$ which starts at $\Sigma_1$
and the cut $\tilde{\Sigma}_2$ which starts at $\Sigma_2$ end at different geodesics,
$\tilde{\cal B}_1$
and $\tilde{\cal B}_2$, respectively.
Any cut which starts at $\Sigma_2$ and ends at ${\cal B}_1$ has to additionally cross the
surface of the torus\footnote{It is easy to understand that this happens because $\Sigma$ cannot
be contracted inside the torus to a point.}.
Such cuts violate the boundary conditions and are prohibited.

\section{Entanglement entropy in brane worlds}
\setcounter{equation}0

Consider now the braneworld models  \cite{RS} where a $d$-dimensional gravity is induced on
a brane embedded in a $(d+1)$-dimensional AdS spacetime. The (Euclidean) RS model has the following
action\footnote{We consider the model with a single brane which
separates two copies  of the AdS spaces.
The copies are cut at the location of the brane ${\cal M}=\partial \tilde{\cal M}$
and glued together. It is implied that
the integrals in (\ref{3.1}) go over the two copies of AdS spaces.}
$$
I_{\mbox{\tiny{RS}}}=-{1 \over 16\pi G_N^{(d+1)}}\int_{\tilde{\cal M}} d^{d+1}x\sqrt{g}
\left(R+{d(d-1) \over l^2}\right)
$$
\begin{equation}\label{3.1}
-{1 \over 8\pi G_N^{(d+1)}}\int_{\partial \tilde{\cal M}} d^{d}x\sqrt{h}K+
I_{\mbox{\tiny{brane}}}~~~.
\end{equation}
Here $I_{\mbox{\tiny{brane}}}$ is the action of the brane. It consists of a contribution
depending on the brane tension $\mu=(d-1)/(4\pi G_N^{(d+1)}l)$ and
a contribution $I_{\mbox{\tiny{matter}}}$ of matter fields localized
on the brane.

The action (\ref{3.1}) is finite because the brane is located at
a finite radius $r$. Conformal infinity of AdS corresponds to the limit
$r\rightarrow 0$.
At small  $r$  the functional
$I_{\mbox{\tiny{RS}}}$ takes the following form (provided that the bulk metric obeys
(\ref{2.10})):
\begin{equation}\label{3.2}
I_{\mbox{\tiny{RS}}}=I_{\mbox{\tiny{matter}}}+\Gamma_{\mbox{\tiny{RS}}}+I_{(1)}+I_{(2)}+...~~~.
\end{equation}
The action is reduced to a non-local part $\Gamma_{\mbox{\tiny{RS}}}$ and a series of
 terms $I_{(k)}$ which are integrals of
$k$-th power polynomials of the curvatures of $\cal M$ and their derivatives.
The first two terms are \cite{ct1}, \cite{ct2}--\cite{ct4}
\begin{equation}\label{3.3}
I_{(1)}=-{1 \over 16\pi G_N^{(d)}}\int_{{\cal M}} d^dx\sqrt{h}R~~~,
\end{equation}
\begin{equation}\label{3.4}
I_{(2)}=-{l^3 \over 16\pi G_N^{(d+1)} (d-2)^2(d-4)}\int_{{\cal M}}
d^dx\sqrt{h}\left(R^{\mu\nu}R_{\mu\nu}
-{d \over 4(d-1)}R^2\right)~~~,
\end{equation}
where $R$ and $R_{\mu\nu}$ are the Ricci scalar and the Ricci tensor of the boundary metric,
and
\begin{equation}\label{3.5}
G_N^{(d)}={(d-2) G_N^{(d+1)} \over 2l}~~~.
\end{equation}
It is implied in (\ref{3.4}) that $d>4$. In $d=4$ one has to replace the factor $1/(d-4)$
in $I_{(2)}$ by $\ln (r /r_0)$, where $r_0$ is  some renormalization
length scale.

From the point of view of the AdS/CFT correspondence the non-local piece,
$\Gamma_{\mbox{\tiny{RS}}}$, can be interpreted as a "renormalized" part of the effective
action for a CFT on the brane
\cite{Gubser},
while local terms $I_{(k)}$ are associated to the ultraviolet diveregences regularized in the
presence of the cutoff $r$. Therefore, the brane theory (in the absence of matter on the brane,
$I_{\mbox{\tiny{matter}}}=0$) is a RS CFT plus gravity theory.
The gravity on the brane is an induced phenomenon, in a sense close to the Sakharov's induced gravity
scenario \cite{Sakh}.

The brane gravitational coupling $G_N^{(d)}$ can be expressed in terms
of the microscopical parameters of the CFT theory. If $d=4$ the CFT is a $U(N)$ superconformal
field theory. The CFT parameters and the supergravity parameters are related by
\begin{equation}\label{g5}
{l^3 \over G_N^{(5)}}={2N^2 \over \pi}~~~.
\end{equation}

To compute the entanglement entropy in the theory on the brane one can proceed as
in Section 2. Let us suppose that geometry is ${\cal M}=\Sigma\times S^1$ where $S^1$ has
circumference length $T^{-1}$. We are interested in the entanglement entropy $S^{RS}(T)$
associated with a part $\Sigma_1$ of $\Sigma$. The entropy can be computed
with the help of the formula
\begin{equation}\label{3.6}
S^{RS}(T)=-\lim_{\beta \rightarrow 2\pi}~ \left(\beta {\partial \over
\partial \beta}-1\right) \ln Z^{RS}(\beta,T)~~~.
\end{equation}
At $\beta=2\pi n$ quantity $Z^{RS}(\beta,T)$ is the partition function of the brane theory
on a manifold ${\cal M}_n$
which is obtained by gluing $n$ copies of ${\cal M}$ along the cut made
along $\Sigma_1$. By using AdS/CFT interpretation of the RS theory one can identify
$Z^{RS}(\beta,T)$ to the path integral in the bulk AdS gravity with the
condition that ${\cal M}_n$ coincides with the boundary
hypersurface embedded in the bulk manifolds $\tilde{\cal M}_n$ at a finite radius. In the semiclassical approximation
\begin{equation}\label{3.7}
\ln Z^{RS}(\beta,T)\simeq -I_{\mbox{\tiny{RS}}}(\beta,T)~~~.
\end{equation}
The classical action, $I_{\mbox{\tiny{RS}}}$,
has the structure (\ref{2.7}) where the bulk metric of $\tilde{\cal M}_n$
solves (\ref{2.10}) and
the induced boundary metric  solves the boundary
Israel equations,
\begin{equation}\label{2.10a}
-{1 \over 4\pi G_N^{(d+1)}}(K_{\mu\nu}-h_{\mu\nu}K)=\mu h_{\mu\nu}+T_{\mu\nu}~~~,
\end{equation}
where $T_{\mu\nu}$ is the stress energy tensor of matter fields on the brane.
The bulk and boundary equations are defined outside the location of conical singularities
where  ${\cal M}_n$ locally coincides with ${\cal M}$. Thus, these equations are the
same as equations in the standard RS setting.
The variational procedure in presence of conical singularities requires that the
singularities in the each copy of bulk AdS
are on a minimal hypersurface $\tilde{\cal B}$ obeying
(\ref{2.11}). The intersection of $\tilde{\cal B}$ with the boundary is the hypersurface
$\cal B$ which divides $\Sigma$ to $\Sigma_1$ and $\Sigma_2$. Therefore, in the semiclassical
approximation
\begin{equation}\label{3.8}
S^{RS}(T)\simeq {{\tilde{\cal A}} \over 4G_N^{(d+1)}}~~~,
\end{equation}
where ${\tilde{\cal A}}$ is the doubled volume of $\tilde{\cal B}$. By using
representation of
the RS action
(\ref{3.2})  one can also write the entropy in another form
\begin{equation}\label{3.9}
S^{RS}(T)=S_{\mbox{\tiny{TH}}}(T)+{{\cal A} \over 4G_N^{(d)}}+S_{(2)}+...
\end{equation}
Here ${\cal A}$ is the volume of $\cal B$ and
$S_{\mbox{\tiny{TH}}}(T)$ is the entanglement entropy in the "renormalized" RS CFT,
\begin{equation}\label{3.10}
S_{\mbox{\tiny{TH}}}(T)=-\lim_{\beta \rightarrow 2\pi}~ \left(\beta {\partial \over
\partial \beta}-1\right) \ln \Gamma_{\mbox{\tiny{RS}}}(\beta,T)~~~.
\end{equation}
The term in (\ref{3.9}) proportional to ${\cal A}$ appears from  $I_{(1)}$.
Corrections $S_{(k)}$ in (\ref{3.9}) correspond to  $I_{(k)}$ with
$k\geq 1$.

Note that the only piece in the entropy (\ref{3.9}) which depends both on the quantum state and
on the hypersurface $\Sigma_1$, where the entangled states are located, is
a "thermal part" $S_{\mbox{\tiny{TH}}}(T)$.
Other terms in $S^{RS}(T)$ have a pure geometrical structure and can be interpreted as a
"vacuum part". The "vacuum parts" of the entropies in $\Sigma_1$ and in its completion
$\Sigma_2$  coincide.

The thermal part $S_{\mbox{\tiny{TH}}}(T)$ dominates over the vacuum part in the high
temperature regime.
It can be shown \cite{RT:06a},\cite{RT:06b} by using properties of minimal surfaces in the AdS
that in this limit $S_{\mbox{\tiny{TH}}}(T)$ is proportional to the volume of $\Sigma_1$,
in agreement with the
extensive properties of the entanglement entropy in quantum field theories at high temperatures
\cite{DF:06a}.

\section{Higher order corrections and conformal anomaly}
\setcounter{equation}0

Let us dwell  on the structure of terms $S_{(k)}$
in (\ref{3.10}) which are determined by functionals $I_{(k)}$,
\begin{equation}\label{4.1}
S_{(k)}=\lim_{\beta \rightarrow 2\pi}~ \left(\beta {\partial \over
\partial \beta}-1\right)I_{(k)}~~.
\end{equation}
The first term $I_{(1)}$ yields $S_{(1)}={\cal A} / (4G_N^{(d)})$, see (\ref{3.9}).
When the separating hypersurface $\cal B$ is a smooth closed manifold
isometrically embedded in ${\cal M}$ and
extrinsic curvatures of $\cal B$ vanish one gets (for $d>4$)
\begin{equation}\label{3.12}
S_{(2)}={l^3 \over 4 G_N^{(d+1)}(d-2)^2(d-4)}\int_{\cal B} d^{d-2} x\sqrt{\sigma}
\left({\cal R}_{ii} -{d \over 2(d-1)}
({\cal R}+2{\cal R}_{ii}-{\cal R}_{ijij})\right)~~~,
\end{equation}
where summation over indexes $i,j=1,2$ is implied.
The quantity $\cal R$ is the Ricci scalar of $\cal B$.
Other two invariants are defined in terms of the Ricci and Riemann tensors of $\cal B$,
${\cal R}_{ii}=n_i^\mu n_i^\nu{\cal R}_{\mu\nu}$,
${\cal R}_{ijij}=n_i^\mu n_j^\nu n_i^\lambda n_j^\rho{\cal R}_{\mu \lambda \nu \rho}$, where
$n_i^\mu$ are two unit vectors in ${\cal M}$ orthogonal to $\cal B$ and normalized,
$(n_i,n_j)=\delta_{ij}$.

Derivation of (\ref{3.12}) is as follows. One keeps in mind that
expressions (\ref{3.2})--(\ref{3.4}) are valid when  the RS action is taken
on a smooth solution of the bulk, (\ref{2.10}), and boundary, (\ref{2.10a}), equations.
Manifolds ${\cal M}_n$  which appear under computation of
the RS partition function $Z^{RS}(\beta,T)$ have conical singularities and
cannot be considered.
Let us replace ${\cal M}_n$ by a smooth manifold which differs
from ${\cal M}_n$ only in some narrow domain around $\cal B$ where
conical singularities are "regularized" by some method (smoothing of conical
singularities is described in \cite{FS1}). The "regularized" bulk manifold
$\tilde{\cal M}_n$
is defined as a solution to bulk equations (\ref{2.10}) with the condition that its
boundary is the regularized ${\cal M}_n$. When regularization  is removed
the brane geometry  approaches ${\cal M}_n$ while bulk space tends to $\tilde{\cal M}_n$.
The RS action on the smoothed manifolds has the asymptotic form
(\ref{3.2}) where $I_{(k)}$ are well defined. As was explained in \cite{FS1},
if one removes the regularization and then takes the limit $\beta=2\pi$ in (\ref{4.1}) the
quantities $S_{(k)}$ will be finite and will not depend on the regularization
procedure\footnote{Note that powers of the curvature are not well-defined when
the regularization is removed. That is why one cannot use this method in deriving
from (\ref{3.2}) gravity theory induced on the brane
with conical singularities.}. In particular,
the results of \cite{FS1} for the polynomials quadratic in curvatures can be used to get
(\ref{3.12}).

A special interest is the entropy in four dimensions, $d=4$. In this case (\ref{3.12})
has to be modified as follows
\begin{equation}\label{3.13}
S_{(2)}=-{l^3 \over 48G_N^{(5)}}~\ln {r \over r_0}~
\int_{\cal B} d^{2} x\sqrt{\sigma}\left(2{\cal R}+{\cal R}_{ii}-
2{\cal R}_{ijij}\right)~~~,
\end{equation}
where the combination $l^3/G_N^{(5)}$ is given in (\ref{g5}). Because of the logarithmic factor
this term breaks the scaling invariance of the CFT.
Explicit connection of $r~ dS_{(2)}/dr$
to the trace anomaly of the stress energy tensor has been found
in \cite{RT:06b}. The structure of the
anomalous piece discussed in \cite{RT:06b} coincides with (\ref{3.13}).

Till now we assumed that the separation surface $\cal B$ has vanishing extrinsic
curvatures. This happens in the case when ${\cal M}$ has a Killing vector field
and $\cal B$ is the set of fixed points of this field. It is the case where results of
\cite{FS1} are applied. The contribution of extrinsic curvatures in $S_{(2)}$
can be established in $d=4$ by using arguments pointed out in \cite{Dowker:94b}.
Because the integral of the anomalous trace of the stress energy tensor is invariant under
the conformal transformations $S_{(2)}$ in this dimension should have the same property.
One has to focus on the transformation of two last terms in the r.h.s. of (\ref{3.13}).
The changes of the metric $h_{\mu\nu} \rightarrow e^{-2\omega} h_{\mu\nu}$
result in the change
\begin{equation}\label{3.14}
{\cal R}_{ii}-2{\cal R}_{ijij}~~\rightarrow ~~{\cal R}_{ii}-2{\cal R}_{ijij}-
2n_i^\mu n_i^\nu(\nabla_\mu\nabla_\nu\omega+\nabla_\mu\omega\nabla_\nu\omega)~~~,
\end{equation}
On the other hand, the extrinsic curvatures transform as
$$
k^i_{\mu\nu}~~\rightarrow ~~e^{-\omega}(k^i_{\mu\nu}+
(n^i)^\lambda \omega_{,\lambda}\sigma_{\mu\nu})~~~,
$$
where $\sigma_{\mu\nu}=h_{\mu\nu}-n^i_\mu n^i_\nu$.
Therefore, the conformally invariant generalization of (\ref{3.13}) in the
presence of non-zero extrinsic curvatures
should be \cite{Dowker:94b}
\begin{equation}\label{3.13a}
S_{(2)}=-{l^3 \over 48G_N^{(5)}}~\ln {r \over r_0}~
\int_{\cal B} d^{2} x\sqrt{\sigma}\left[2{\cal R}+{\cal R}_{ii}-
2{\cal R}_{ijij}+\frac 12 (k^i)^2+\lambda\left((k^i)^2-2 \mbox{Tr}(k^ik^i)\right)\right]~~~,
\end{equation}
where $k^i=(k^i)^\mu_\mu$, $\mbox{Tr}(k^ik^i)=(k^i)_{\mu\nu}(k^i)^{\mu\nu}$.
The combination $(k^i)^2-2 \mbox{Tr}(k^ik^i)$ is conformally invariant  and the
constant $\lambda$ has to be determined by a different method.

Let us also note  that $\cal B$ was assumed to be a smooth closed manifold in $\Sigma$.
If $\cal B$ has boundaries there will be extra contributions to $S_{(k)}$.
Calculations of the entanglement entropy in QFT's taking into account boundary
effects can be found in \cite{DF:06a}.

\section{Effect of higher curvature terms in the bulk}
\setcounter{equation}0

Quantum effects in the bulk AdS space may modify holographic formula (\ref{1.3}).
One type of these modifications is the appearance of
higher curvature terms in the bulk action (\ref{2.2}).
Let us consider a simple example when the higher curvature term added to the r.h.s. of (\ref{2.2})
is the Gauss-Bonnet (GB) term \footnote{The gravitational action with the GB term
appears in the low-energy string models \cite{Zw}.},
\begin{equation}\label{5.1}
I_{GB}=-{\alpha \over 16\pi G_N^{(d+1)}}\int_{\tilde{\cal M}} d^{d+1}x\sqrt{g}
\left(R^2-4R_{KL}R^{KL}+R_{KLMN}R^{KLMN}\right)+I_{GB,b}~~~.
\end{equation}
Here $\alpha$ is some coefficient with the dimension of length square and $I_{GB,b}$
is the corresponding boundary term. The derivation of entanglement entropy
in the modified theory is the same as in Section 2.

To find the entropy
one has to consider the bulk action on spaces $\tilde{\cal M}_n$ which have conical
singularities on a codimension 2 hypersurface $\tilde{\cal B}$, a holographic dual of
the corresponding separating surface $\cal B$.
The  bulk gravity action with the GB term can be written as
\begin{equation}\label{5.2}
I_{gr}=I_{gr}'+(\beta-2\pi){\cal I}~~~,
\end{equation}
where $I_{gr}'$ is the action functional on the regular
domain $\tilde{\cal M}_n/\tilde{\cal B}$ of $\tilde{\cal M}_n$ and
\footnote{To find contribution
of the conical singularities in (\ref{5.1}) we used results of \cite{FS1}.
We also assumed for simplicity that extrinsic curvatures of
$\tilde{\cal B}$ vanish. Possible effects of conical singularities
in  $I_{GB,b}$ in (\ref{4.1}) as well as boundary terms in (\ref{5.3}) are ignored.}
\begin{equation}\label{5.3}
{\cal I}={1 \over 8\pi G_N^{(d+1)}}\int_{\tilde{\cal B}}d^{d-1}x\sqrt{\tilde{\sigma}}
\left(1+2\alpha ~\tilde{\cal R}\right)~~~.
\end{equation}
Here $\tilde{\sigma}$ is the determinant of metric tensor $\tilde{\sigma}_{ab}$
of $\tilde{\cal B}$ and $\tilde{\cal R}$ is the Ricci scalar of $\tilde{\cal B}$.
The entanglement entropy in the semiclassical approximation is
\begin{equation}\label{5.4}
S^{CFT}=2\pi {\cal I}={{\tilde{\cal A}} \over 4G_N^{(d+1)}}+
{\alpha \over 2 G_N^{(d+1)}}\int_{\tilde{\cal B}}d^{d-1}x\sqrt{\tilde{\sigma}}~
\tilde{\cal R}~~~
\end{equation}
which is the result of application (\ref{2.5}) to (\ref{5.2}). Thus, the Gauss-Bonnet term
yields a correction to holographic formula (\ref{1.3}) which is proportional to the
integral curvature of $\tilde{\cal B}$. Modification of (\ref{1.3}) by the GB term
has been also discussed in \cite{IKSY}.

The bulk metric $\bar{g}_{KL}$ now has to be a solution to equations (\ref{2.10})
modified by GB terms. Equations which determine position of $\tilde{\cal B}$ correspond
to an extremum of functional (\ref{5.3}), $\delta {\cal I}=0$. Let $\zeta^a$,
$a=1,...,d-1$ be some coordinates on $\tilde{\cal B}$. The embedding of
$\tilde{\cal B}$ is described by equations $X^K=X^K(\zeta)$. Variations of $X^K$ in
$\cal I$ on
the given bulk metric $\bar{g}_{KL}$ yield
\begin{equation}\label{5.5}
T^{ab}\left(X^K_{~~||ab}+\bar{G}^K_{LM}X^L_{~~,a}X^M_{~~,b}\right)
+T^{ab}_{~~~||a}X^K_{~~,b}=0~~~,
\end{equation}
\begin{equation}\label{5.6}
T^{ab}=\sigma^{ab}+4\alpha\left(\frac 12 \sigma^{ab}\tilde{\cal R}-
\tilde{\cal R}^{ab}\right)~~~.
\end{equation}
Here $\sigma^{ab}$ is the inverse matrix of the metric
$\sigma_{ab}=\bar{g}_{KL} X^K_{~~,a}X^L_{~~,b}$ induced
on $\tilde{\cal B}$, symbol $||$ denotes covariant derivatives on  $\tilde{\cal B}$,
and $\bar{G}^K_{LM}$ are the connections for the bulk metric.
If $\alpha=0$ equations (\ref{5.5}) reduce to the standard Nambu-Goto equations.

\section{Discussion}
\setcounter{equation}0

The main purpose of our work was to give a proof of the holographic formula (\ref{1.3})
for the entanglement entropy in QFT's which have a dual description in terms of the
AdS gravity. In Section 2 we have formulated conditions under which (\ref{1.3})
can be applied. In particular, we demonstrated that $\tilde{\cal B}$, a holographic dual
of the separating surface $\cal B$, has to be embedded
in a Riemannian AdS space. We also pointed out topological obstructions on the choice of
$\tilde{\cal B}$. Our analysis incorporates static problems discussed
in \cite{RT:06a}, \cite{RT:06b}, as well
as the case of static black holes on the brane when the separating surface is associated
to the black hole horizon \cite{Em:06}. It also allows one to extend computations in stationary
but non-static theories along the lines explained in Section 2.

Let us return to formula (\ref{1.1}) and compare it with result (\ref{3.9}) obtained
in the RS braneworld model. Suppose that the braneworld geometry is flat, ${\cal M}=R^d$.
Suppose also that the system is in the ground state and the separating surface
is a plane. In this case terms $S_{\mbox{\tiny{TH}}}(T)$ and $S_{(k)}$ for $k\geq 2$
vanish ($S_{(k)}$ are some invariants
constructed in terms  of extrinsic and intrinsic
geometrical characteristics of $\cal B$).
The density of entanglement entropy ${\cal S}^{RS}$ per unit area of $\cal B$ becomes
\begin{equation}\label{3.11}
{\cal S}^{RS}= {1 \over 4G_N^{(d)}}~~~,
\end{equation}
in agreement with (\ref{1.1}).
In general there should be corrections to (\ref{3.11}) when
${\cal M}$ and $\cal B$ are not flat.

According to (\ref{3.11}) quantum entanglement of the degrees of freedom of
the   fundamental gravity  theory
can be measured in terms of the gravitational coupling $G_N^{(d)}$
on the brane. There are two ways to describe these degrees of freedom.
At energies below the scale $l^{-1}$ these degrees of
freedom are the field variables of the brane CFT with the UV cutoff at $l^{-1}$.
In four dimensions the gravity induced by such a CFT has the coupling
$G_N^{(d)}\sim l^2/N^2$, which agrees with Eq. (\ref{3.5}).
At the energies above $l^{-1}$ the brane CFT is itself an effective theory. The genuine
degrees of freedom are related to the string theory which provides an exact
expression $G_N^{(4)}=\pi l^2/(2N^2)$, where $l=\lambda^{1/4} l_s$,
$\lambda=g_{YM}^2N$ and $l_s$
is the string length.

We do not give a direct microscopical counting of the
entanglement entropy on the brane. Our analysis is suggestive.
However, the described picture  is very similar
to that in condensed matter models.
Consider the ground state entanglement entropy $S$ in the spin chains near
the critical point \cite{Rico}, for instance, in the Ising model. Here the
entropy can be directly derived from the density matrix
determined by the Ising Hamiltonian. Near the critical point the Ising model
corresponds to a 2D QFT with two massive fermion fields. This QFT at the critical point
becomes a conformal theory.
Thus, the alternative derivation of $S$ can be given in terms of the effective QFT
with the UV cutoff associated with the inverse lattice spacing. The two derivations
yield the same result for $S$, see the details in \cite{Rico}.

\bigskip

\noindent
\section*{Acknowledgment}\noindent
I am grateful D.V. Vassilevich for useful comments.
This work was supported by the Scientific School Grant N 5332.2006.2.

\end{document}